\documentclass{eptcs}
\usepackage{graphicx}
\usepackage{color}
\definecolor{lstbg}{rgb}{0.98,0.98,0.98}

\usepackage{listings}
\lstdefinelanguage{gretl}{
  morekeywords={E, V, as, bag, eSubgraph, end, exists!, exists, false, forall,
    from, import, in, let, list, null, path, pathSystem, rec, report,
    reportBag, reportMap, reportSet, thisEdge, thisVertex, map,
    role,    
    set, store,
    true, tup, using, vSubgraph, where, with, to},
  morekeywords={new, protected, public, static, final, private, if, switch,
    case, for, break, default, throw, void, else, return, continue},
  morekeywords={[2]and, avg, contains, containsKey, count, degree, depth,
    difference, distance, dividedBy, edgesConnected, edgesFrom, edgesTo,
    edgeTrace, edgeTypeSet, endVertex, enumConstant, equals, extractPath,
    getEdge, get, getValue, getVertex, grEqual, grThan, hasAttribute, hasType,
    id, inDegree, innerNodes, intersection, isAcyclic, isEmpty, isA, isCycle,
    isIn, isIsolated, isLoop, isNeighbour, isNull, isParallel, isReachable,
    isSibling, isSubPathOfPath, isSubSet, isSuperSet, isTrail, isTree, keySet,
    leaves, leEqual, leThan, maxPathLength, minPathLength, minus,
    modulo, nequals, nodeTrace, not, nthElement, theElement, or, outDegree,
    package-info, parent, pathConcat, pathLength, pathSystem, plus, pos,
    reachableVertices, recordInstance, reMatch, schemaFunctions, siblings,
    squareRoot, startVertex, subtypes, sum, supertypes, symDifference, times,
    toString, type, typeName, typeSet, types, uminus, union, values,
    vertexTypeSet, weight, xor},
  emph={AddMappings, CreateSubgraph, CreateVertexClass,
    CreateAbstractVertexClass, CreateEdgeClass, CreateAbstractEdgeClass,
    AddSubClass, AddSubClasses, AddSuperClass, AddSuperClasses, CopyDomain,
    CreateAttribute, CreateAttributes, CreateListDomain, CreateSetDomain,
    CreateMapDomain, CreateRecordDomain, CreateEnumDomain, CreateVertices,
    CreateEdges, MatchReplace, MergeVertices, Delete, SetAttributes,
    RedefineFromRole, RedefineFromRoles, RedefineToRole, RedefineToRoles,
    transformation, Iteratively},
  emphstyle=\bf\underbar,
  sensitive=true,
  string=[b]{\\"},
  morecomment=[l]{//}}

\makeatletter
\lst@AddToHook{TextStyle}{\let\lst@basicstyle\normalsize\sffamily}
\makeatother

\newcommand{\myemail}[0]{\email{horn@uni-koblenz.de}}
\newcommand{\gretl}[1]{\lstinputlisting[language=gretl]{#1}}

\def\titlerunning{Saying Hello World with GReTL -- A Solution to the TTC 2011 Instructive Case}
\def\authorrunning{Tassilo Horn}

\title{Saying Hello World with GReTL -- A Solution to the TTC 2011 Instructive Case}
\author{Dipl.-Inform. Tassilo Horn\\
  \myemail\\
  Institute for Software Technology\\
  University Koblenz-Landau, Campus Koblenz}

\begin{document}
\maketitle
\begin{abstract}
  This paper discusses the GReTL solution of the TTC 2011 Hello World case
  \cite{helloworldcase}.  The submitted solution covers all tasks including the
  optional ones.
\end{abstract}

\section{Introduction}
\label{sec:introduction}

GReTL (Graph Repository Transformation Language, \cite{gretl-icmt2011}) is the
operational transformation language of the TGraph technological space
\cite{tgapproach08}.  Models are represented as typed, directed, ordered, and
attributed graphs.  There are import/export tools for EMF models and
metamodels.  GReTL is designed as a Java API, but a simle concrete syntax is
provided as well.

The elementary GReTL transformation operations follow the conception of
incrementally constructing the target metamodel together with the target graph.
In this challenge's tasks, the target metamodels are provided, so only
operations working on the instance level are used, i.e., the operations create
new vertices and edges in the target graph, and they set attribute values.
GReQL queries \cite{FestschriftNagl2010} evaluated on the source graph are used
to specify what has to be created.

Besides these elementary out-place operations, GReTL offers a set of in-place
operations that allow for deleting elements or replacing elements matched by a
pattern with some subgraph similar to rules of graph replacement systems.  In
the next section, all mandatory Hello Word tasks are discussed, thereby
explaining GReTL in some details.  The solutions of the optional tasks are
discussed in the appendix.

\section{Task Solutions}
\label{sec:task-solutions}

\sloppy{In this section, all mandatory tasks are discussed in sequence and the
  GReTL transformations and GReQL queries are explained when they come along.
  The solution can be run on SHARE \cite{share-demo}.

\paragraph{Task 1: Hello World.}
The first task is to create one single \lstinline{Greeting} vertex.

\gretl{1-hello-world.gretl}

Line 1 declares the name of the transformation.  In line 2, the
\lstinline{CreateVertices} operation is invoked.  The argument specifies the
type of the vertices to be created, i.e., \lstinline{Greeting}.  After the
arrow symbol, there is a GReQL query that is evaluated on the source graph and
should return an arbitrary set.  For any member of that set, a new
\lstinline{Greeting} vertex is be created, and the mapping from set member to
new vertex is saved and can be used in following operation calls.

Since this is a constant transformation, there is no source graph.  The query
evaluates to a set containing only the number 1.  Thus, one new
\lstinline{Greeting} vertex is created.  The mapping from 1 to that vertex is
saved in a function corresponding to the target element type
(\lstinline{img_Greeting}).  In this context, 1 is called the \emph{archetype}
of the new \lstinline{Greeting} vertex.  The new \textsf{Greeting} vertex is in
turn called the \emph{image} of 1.

In line 3, the \lstinline{text} attribute of \lstinline{Greeting} vertices is
set.  The \lstinline{SetAttributes} operation expects a map which assigns to
each archetype of some vertex or edge the value that its corresponding target
graph image should have set for the specified attribute.  In this case, the map
contains only one single entry: the integer 1 maps to the string ``Hello
World''.  Thus, for the image of the number 1 in \lstinline{img_Greeting},
i.e., the new \lstinline{Greeting} vertex, the \lstinline{text} attribute is
set to ``Hello World''.

\paragraph{Task 2: Hello World Extended.}

The second task is to create an extended hello world graph.

\gretl{2-hello-world.gretl}

The \lstinline{CreateSubgraph} operation is similar to
\lstinline{CreateVertices} in that it gets a GReQL query resulting in a set.
For each member in that set, a subgraph specified by the template preceeding
the \lstinline{<==} symbol is created.  Parenthesized constructs denote
vertices to be created, and arrow symbols with curly braces denote edges to be
created.  In each template vertex or edge, first the type name is given, then
an archetype, followed by an optional list of attribute-value pairs.  For
edges, the archetypes may be omitted, but for vertices the archetype is
mandatory, because it is used internally to refer to the new vertices when
creating the edges connecting them.  The \$ variable refers to the current
member of the set returned by the query.  Since the set contains only one
member, the template will be evaluated only once, and the single binding of \$
is 1.  Three vertices (a \textsf{GreetingMessage}, a \textsf{Greeting}, and a
\textsf{Person}) and two connecting edges (a
\textsf{GreetingContainsGreetingMessage} edge, and a
\textsf{GreetingContainsPerson} edge) are created.  The archetypes of all
vertices is the value of \$, i.e., 1.

\paragraph{Task 3: Model-2-Text.}

A GReQL query is used to serialize the graph created in the last task.

\gretl{3-hello-to-text.greql}

For any \lstinline{Greeting} vertex, a string is created by concatenating the
\lstinline{text} of that greeting's \lstinline{GreetingMessage}, one space, the
\lstinline{name} of that greeting's \lstinline{Person}, and finally an
exclamation mark.  The result is a set of strings.  Since the graph contains
only one greeting, it is a set with one single string ``Hello TTC
Participants!''.  The expression \lstinline|greet <>--{GreetingContainsPerson}|
calculates the set of vertices reachable from \textsf{greet} by traversing a
containment edge of type \textsf{GreetingContainsPerson} in the direction from
part to whole.  Because there must be exactly one person associated to a
greeting by such an edge, the function \textsf{theElement()} is used to extract
it.

\paragraph{Task 4: Count Nodes.}

In this task, the number of \lstinline{Node} vertices is to be determined.

\gretl{4-count-nodes.greql}

\paragraph{Task 5: Count Looping Edges.}

Note that because all models were imported from EMF without any optimization,
in the TGraph all elements of type \lstinline{Edge_}\footnote{The underscore
  has been appended because \textsf{Edge} is the abstract base type of all edge
  types and thus a reserved word.} are in fact vertices, and the
\lstinline{src} and \lstinline{trg} references are real edges of the types
\lstinline{Edge_LinksToSrc} and \lstinline{Edge_LinksToTrg}.

\gretl{5-count-loops.greql}

First, all loops are determined by selecting those \lstinline{Edge_} vertices
whose \lstinline{src} and \lstinline{trg} edges point to the same vertex.  A
tuple is returned that contains the number of loops and the set of loops.  This
is not required by the task, but it should be noted that the result tuple could
be processed in Java similar to a SQL result set.

\paragraph{Task 6: Isolated Nodes.}

To find isolated nodes, the following GReQL query is used:

\gretl{6-isolated-nodes.greql}

First, all isolated nodes are determined by restricting the nodes to those
which are not connected to any \lstinline{Edge_LinksToSrc} or
\lstinline{Edge_LinksToTrg} edge.  Instead of the nodes themselves, the names
are selected for a better comparison with the EMF models.  As result a string
is constructed that mentions the number of isolated nodes and lists them.

\paragraph{Task 7: Circle of Three Nodes.}

The following GReQL query is used:

\gretl{7-circe-of-three.greql}

The variables \lstinline{n1}, \lstinline{n2}, and \lstinline{n3} iterate over
all \lstinline{Node} vertices.  The \lstinline{with}-part ensures they are
pairwise distinct and form a circle.  For example, between \textsf{n1} and
\textsf{n2} there has to be a vertex of type \textsf{Edge\_} which references
\textsf{n1} and \textsf{n2}.  The query results in a string mentioning the
number of circles and lists them.

\paragraph{Task 9: Reverse Edges.}
\label{sec:reverse-edges}

The task of reversing edges is done using a GReTL in-place transformation.
Here, the operation \lstinline{MatchReplace} is used.  Just like the
\textsf{CreateSubgraph} operation used in Task 2, it receives a template graph.
Analogously, it receives a GReQL query following the \lstinline{<==} symbol.
The query results in a set, and for each member in the set (a match), the
template graph is applied.  The elements in the template graph may refer to
things in the current current match via the variable \$.  All elements in a
match that are used in the template graph are preserved, all elements in a
match that are not used are deleted, and elements of the template graph that
don't reference an element in the match are created.

\gretl{9-reverse-edges.gretl}

The query reports a set of 5-tuples, one tuple for each \lstinline{Edge_}
vertex, including its source and target nodes, and the corresponding
\textsf{Edge\_LinksToSrc} and \textsf{Edge\_LinksToTrg} edge.

For each match, the template graph is applied with the current match tuple
bound to \$.  \textsf{\$[0]} references the match's \lstinline{Edge_} vertex by
its index in the reported match tuple using an array-like syntax.  Likewise,
\textsf{\$[1]} references the source node, and \textsf{\$[2]} references the
target node.  Since these nodes are referenced, they are preserved.  The edges
in each match tuple (\textsf{\$[3]} and \textsf{\$[4]}) are not referenced in
the template and thus deleted.  Two new \textsf{Edge\_LinksToTrg} and
\textsf{Edge\_LinksToSrc} edges are created, but this time the former source
node is the target and the former target node is the source.

\paragraph{Task 10: Simple Migration.}

This task is solved with an out-place GReTL transformation.  In lines 2 to 4,
the vertices of type \lstinline{Graph_}, \lstinline{Node}, and
\lstinline{Edge_} are ``copied'' over, i.e., for any \textsf{Graph\_} node in
the source graph, a \textsf{Graph\_} node in the target graph is created and
likewise for \textsf{Node} and \textsf{Edge\_} nodes.

\gretl{10-simple-migration.gretl}

Then the \lstinline{GraphComponent.text} attribute is set in lines 5 to 7.  The
query returns a map that assigns to each \lstinline{GraphComponent} archetype,
i.e., a source graph \lstinline{Graph_}, \lstinline{Node}, or \lstinline{Edge}
vertex, the value that its image in the target graph should have set for the
\lstinline{text} attribute.  If the archetype is of type \lstinline{Node}, then
the value is the contents of its \lstinline{name} attribute.  Else, it is the
empty string.

From line 8 on, the edges are ``copied'' into the target graph.  For any source
graph \lstinline{Edge_LinksToSrc} edge a target graph
\lstinline{Edge_LinksToSrc} edge is created.  Because edges cannot exist on
their own, the query given to \textsf{CreateEdges} has to result in a set of
triples.  The first component is the archetype of the new edge which can be
used in following operations to refer to it.  The second and third component
are the archetype of the new edge's start and end vertex.  Thus, the new target
graph \lstinline{Edge_LinksToSrc} edge starts at the image of its source
counterpart's start vertex and it ends at the image of the end vertex.

In lines 12 to 14, the \lstinline{Graph_ContainsGcs} edges are created.  Each
of those corresponds to either a source graph \lstinline{Graph_ContainsNodes}
or a \lstinline{Graph_ConstainsEdges} edge.

\paragraph{Task 12: Delete Node n1.}

In this task, all nodes with \lstinline{name} attribute set to ``n1'' should be
removed.

\gretl{12-delete-node-n1.gretl}

The \lstinline{Delete} operation deletes all elements returned by the query.

\paragraph{Task 13: Delete Node n1 and Connected Edges.}

This task is similar to the previous task, except that all \lstinline{Edge_}
vertices connected to the \lstinline{Node} to be deleted should be deleted,
too.

\gretl{13-delete-node-n1-and-inc-edges.gretl}

So the query returns the nodes to be deleted and all \lstinline{Edge_} vertices
reachable by traversing edges targeting \lstinline{n} where \lstinline{n} is in
the \lstinline{src} or \lstinline{trg} role.

\section{Conclusion}
\label{sec:conclusion}

In this paper, the solutions of all Hello World tasks have been briefly
discussed.  Most transformations were implemented using the elementary GReTL
operations \textsf{CreateVertices}, \textsf{CreateEdges}, and
\textsf{SetAttributes} whose concepts are explained in more details in
\cite{gretl-icmt2011}.  GReTL is a very extensible language, and the in-place
operations \textsf{MatchReplace}, \textsf{Iteratively}, and \textsf{Delete}
used in some task solutions were added to the language for solving the
\emph{Compiler Optimization} case
\cite{compileroptimizationcase,compileroptimizationsolutiongretl}.

\bibliographystyle{eptcs}
\bibliography{bibliography}

\clearpage
\appendix

\section{Appendix: Optional Tasks}
\label{sec:append-opti-tasks}

In this short appendix, the three optional tasks are discussed.

\paragraph{Optional Task 8: Dangling Edges.}

Here, ``dangling edge'' refers to a vertex of type \lstinline{Edge_} which has
only one outgoing edge.

\gretl{8-dangling-edges.greql}

The vertices of type \lstinline{Edge_} are restricted to those that have
exactly one incident edge of type \lstinline{Edge_LinksToSrc} or
\lstinline{Edge_LinksToTrg}.  Those are counted and listed.

\paragraph{Optional Task 11: Topology Changing.}

In this task, the conceptual edges represented as vertices should be
transformed into real edges.  The transformation is quite similar to the last
one.  The \lstinline{Graph_} and \lstinline{Node} vertices but no
\lstinline{Edge_} vertices are created in the target graph, the
\lstinline{text} attributes of nodes are set, and the
\lstinline{Graph_ContainsNodes} edges are created.

\gretl{11-topology-chaining.gretl}

The interesting operation is the last one, which creates the
\lstinline{NodeLinksToLinksTo} edges (the strange name is generated by our
Ecore importer from the class and role names in the Ecore metamodel).  For each
source graph \lstinline{Edge_} vertex \lstinline{e}, the query reports a triple
containing \lstinline{e}, the \lstinline{Node} vertex at the \lstinline{src}
end of the outgoing \lstinline{Edge_LinksToSrc} edge, and the \lstinline{Node}
vertex at the \lstinline{trg} end of the \lstinline{Edge_LinksToTrg} edge
starting at \lstinline{e}.  Thus there will be a new
\textsf{NodeLinksToLinksTo} edge for every \textsf{Edge\_} vertex which
starts/ends at the \textsf{Node} vertices the original \textsf{Edge\_}
referenced.

\paragraph{Optional Task 14: Insert Transitive Edges.}
\label{sec:optional-task-14}

In this task, transitive edges should be created.  It is implemented as
in-place transformation on graphs conforming to the ``topology changing''
metamodel (Fig. 6 in the case description \cite{helloworldcase}).

\gretl{14-transitive-edges.gretl}

This transformation uses a little trick in order to create only transitive
edges of the original graph, but not the transitive closure.  Therefore, the
set of \lstinline{matches} is calculated beforehand.  This is a set of
\textsf{Node} pairs where a transitive edge has to be created in between.

\lstinline{Iteratively} is a higher-order operation that executes the following
transformation operations as long as any of them is applicable.  The query of
the \lstinline{MatchReplace} call iterates over the pairs of nodes and checks
if no transitive edge created by a previous iteration exists.  In that case, the
template graph specifies the creation of such an edge.

The little trick is required for this reason: Although the query provided to
\lstinline{MatchReplace} results in a set of matches, the operation skips
matches containing elements that already occured in previous matches.  These
elements might have been modified in a way that invalidates the current match.

\end{document}